\documentstyle[12pt,epsfig]{article}

\newlength{\dinwidth}
\newlength{\dinmargin}
\begin{document}
\def\bold#1{\setbox0=\hbox{$#1$}%
     \kern-.025em\copy0\kern-\wd0
     \kern.05em\copy0\kern-\wd0
     \kern-.025em\raise.0433em\box0 }
\def\slash#1{\setbox0=\hbox{$#1$}#1\hskip-\wd0\dimen0=5pt\advance
       \dimen0 by-\ht0\advance\dimen0 by\dp0\lower0.5\dimen0\hbox
         to\wd0{\hss\sl/\/\hss}}
\def\lq{\left [}
\def\rq{\right ]}
\def\LL{{\cal L}}
\def\VV{{\cal V}}
\def\AA{{\cal A}}
\def\BB{{\cal B}}
\def\MM{{\cal M}}
\def\ovl{\overline}
   
\newcommand{\be}{\begin{equation}}
\newcommand{\ee}{\end{equation}}
\newcommand{\bea}{\begin{eqnarray}}
\newcommand{\eea}{\end{eqnarray}}
\newcommand{\nn}{\nonumber}
\newcommand{\dd}{\displaystyle}
\newcommand{\bra}[1]{\left\langle #1 \right|}
\newcommand{\ket}[1]{\left| #1 \right\rangle}
\newcommand{\spur}[1]{\not\! #1 \,}
\thispagestyle{empty}
\vspace*{1cm}
\rightline{BARI-TH/97-262}
\rightline{March 1997}
\vspace*{2cm}
\begin{center}
  \begin{LARGE}
  \begin{bf}
\vskip 1cm
On the QCD Sum Rule Determination\\ of the Strange Quark Mass\\
  \end{bf}
  \end{LARGE}
  \vspace{8mm}
  \begin{large}
P. Colangelo$^a$, F. De Fazio$^{a,b}$, G. Nardulli$^{a,b}$, N. Paver$^{c}$
  \end{large}
  \vspace{1cm}

\begin{it}
$^{a}$ Istituto Nazionale di Fisica Nucleare, Sezione di Bari, Italy\\
$^{b}$ Dipartimento di Fisica, Universit\'a di Bari, Italy \\
$^{c}$ Dipartimento di Fisica Teorica, Universit\'a
di Trieste, Italy, and  \\  
Istituto Nazionale di Fisica Nucleare, Sezione di Trieste, Italy\\
\end{it}
\end{center}
\begin{quotation}
\vspace*{1.5cm}
\begin{center}
  \begin{bf}
  Abstract\\
  \end{bf}
\end{center}
\vspace{5mm}
\noindent
In the QCD Sum Rule determination of $m_s$ using the two-point correlator
of divergences of $\Delta S=1$ vector currents, the final uncertainty on $m_s$
is mainly due to the hadronic spectral 
function. Using a specific parameterization which fully takes into account the 
available experimental data on the $K \pi$ ($I=1/2, \; J^P=0^+)$ system, 
characterized by the presence of a relevant nonresonant component in 
addition to the resonant one, we find 
${\overline m_s}(1 \; GeV)\ge 120 \; MeV$. 
In particular, varying only the parameters describing the nonresonant $K \pi$
component and $\Lambda_{\overline {MS}}^{n_f=3}$ we obtain
${\overline m_s}(1 \; GeV)=125 - 160 \; MeV$.
This result is 
smaller than analogous ones obtained by using a parameterization in terms of
only resonant states. We discuss how to systematically improve the 
determination of $m_s$ by this method.

\vspace*{0.5cm}
\end{quotation}

\newpage
\baselineskip=18pt
\setcounter{page}{2}

Light `current' quark masses have an important role in the theoretical 
description of low energy hadronic physics, and their actual values are 
needed as an input to quantitatively predict several relevant effects, such 
as, e.g., the breaking of chiral and vector flavour symmetries in hadron 
masses and transition amplitudes, and the size of CP violation in Kaon decays. 

Chiral perturbation theory represents a consistent framework where the 
(scale-inde\-pendent) ratios 
of `current' quark masses can be evaluated from experimental data.
At the next to leading  order in the chiral expansion one obtains 
\cite{gasser,gasser1}:
\begin{equation}
{ m_s \over  m_d} = 18.9 \pm 0.8 \;\;\; ; \;\;\;\; 
{ m_u \over m_d } = 0.553 \pm 0.043 \;\;\; .\label{cratios}  
\end{equation}

However, for the determination of the individual light quark masses from the 
experimental data, 
one has to adopt some alternative non-perturbative method, such as lattice QCD 
or QCD sum rules. Once (at least) one of these masses is determined from 
these methods, the ratios in Eq.~(\ref{cratios}) give access to the other ones. 
Here, we shall consider the determination of the strange quark mass 
$m_s$ from QCD sum rules. The reason for concentrating on $m_s$ (rather than 
on the other masses) is that there is more experimental 
information concerning the hadronic spectral function on which the 
QCD sum rule method  is based,
 so that in this case one could expect the final result 
to be less model-dependent and more reliable.

The present situation concerning the evaluation of $m_s$ is the 
following. Lattice QCD calculations, in quenched approximation, give 
for the running mass:
\begin{eqnarray}
{\overline m_s}(2 \; GeV) &=& 128 \pm 18 \; MeV  \;\;\;\;\cite{allton}  \\
{\overline  m_s}(2 \; GeV) &=& 100 \pm 21 \pm 10 \; MeV \;\;\;\cite{gupta}\; .  
\label{gupta}
\end{eqnarray}
For $\Lambda_{\overline {MS}}^{n_f=3}=380 \; MeV$ these values correspond to
${\overline  m_s}(1 \; GeV) = 172 \pm 24 \; MeV$, and
${\overline  m_s}(1 \; GeV) = 140 \pm 29 \pm 14 \; MeV$, respectively.

Regarding QCD sum rules, substantial progress has been achieved quite recently  
by the calculation to ${\cal O}(\alpha_s^3)$ of the perturbative part 
of the two-point correlator of $\Delta S=1$ scalar quark currents relevant 
to the determination of $m_s$ \cite{chetyrkin}. 
Accounting for such ${\cal O}(\alpha_s^3)$ corrections, the value previously 
obtained in \cite{jamin} by using the ${\cal O}(\alpha_s^2)$ expression of the 
perturbative contribution to the sum rule:\footnote{ 
\baselineskip 12pt 
Earlier QCD sum rules estimates, using different versions of the method, 
can be found, e.g., in refs. \cite{old_ms,old_ms_1,old_ms_2}.} 
\begin{equation}
{\overline m_s}(1 \; GeV) = 189 \pm 32 \; MeV \;\;\;\cite{jamin} \label{qcd1}
\end{equation}
has changed to the new value:
\begin{equation}
{\overline  m_s}(1 \; GeV) = 203.5 \pm 20 \; MeV \;\;\; 
\cite{chetyrkin} \;\;\;. \label{qcd2}
\end{equation}
The uncertainties in Eqs.~(\ref{qcd1},\ref{qcd2}) are 
related to the dependence of the result
on the value of $\Lambda_{\overline {MS}}^{n_f=3}$, which was chosen in the range 
$280-480\; MeV$, and to the variation of other input parameters in the sum rule 
analysis. In particular, as concluded in \cite{chetyrkin}, the largely 
dominant uncertainty is that on the hadronic spectral function, which  
in both \cite{jamin} and \cite{chetyrkin} is assumed to behave as the sum of 
two $I=1/2, J^P=0^+$  $(K\pi)$ resonant states.

Following this observation, we would like to discuss a 
parameterization of the relevant hadronic spectral function which, at least in 
principle, should be able to fully exploit 
the current experimental information 
on the scalar $I=1/2$ channel. In particular, this parameterization takes 
into account the fact that, 
 in addition to the resonance component, there exists also
a nonresonant continuum.
As typical of $s$-wave channels, the nonresonant component
persists rather far from the 
threshold energy range and can interfere with the resonances \cite{tornqvist}.
Our example shows that the inclusion of the nonresonant contribution may 
give a significant effect on the sum rule and, also, indicates some directions 
in order to improve the determination of $m_s$ in this approach.   

For convenience, we briefly describe the main points of the QCD sum rule 
derivation of $m_s$ we are interested in. The basic quantity is the two-point 
correlator
\begin{equation}
\Psi(q^2)= i \int dx \; e^{i q x} <0| T[J(x) J^\dagger(0)] |0> \label{corr}
\end{equation}
where $J(x)=i (m_s - m_u) {\bar s}(x) u(x)$ is the divergence of the 
$\Delta S =1$ vector current $V^\mu(x)= {\bar s}(x) \gamma ^\mu u(x)$. 
The second derivative 
$\Psi^{''}(q^2)= ( \partial^2/(\partial q^2)^2 ) \Psi(q^2)$ obeys an 
unsubtracted dispersion relation:
 \begin{equation}
\Psi^{''}(q^2) = 2 \int_0^\infty ds {\rho(s) \over (s - q^2 -i \epsilon)^3}\;,
\label{psisec}
\end{equation}
with the spectral function $\rho(s)$ given by
\begin{equation}
\rho(s)= {1 \over \pi} Im \; \Psi(s) \;\;\;.
\end{equation}
Particularly useful 
for low-energy phenomenology is the Borel 
transform $\Psi^{''}(M^2)$ \cite{svz}, defined by the application 
 of the operator 
$\displaystyle B(M^2)={(-1)^n \over n!} \Big({d \over d (-q^2)})^n$ 
in the limit
$\displaystyle n \to \infty$, $-q^2 \to \infty$, 
$\displaystyle -q^2/n=M^2=const$ to Eq.~(\ref{psisec}). 
This results into: 
\begin{equation}
\Psi^{''}(M^2)={1 \over M^6} \int_0^\infty ds \; \rho(s) 
e^{-s/M^2} \;\;\;.\label{borpsi} 
\end{equation}
The main advantage of this transformation is that, due to the presence of the 
exponential, for moderate values of the Borel parameter $M^2$ (of the order 
of one to a few $GeV^2$) the r.h.s of Eq.~(\ref{borpsi}) is mostly sensitive 
to the spectral function $\rho(s)$ in the low energy range, where it can be 
computed in terms of the available experimental information.  

The quark `current' mass $m_s$ enters into the l.h.s 
of Eq.~(\ref{psisec}) as a parameter, when the correlation function 
$\Psi^{''}(Q^2)$ is evaluated in QCD for $Q^2 = -q^2 \gg \Lambda^2_{QCD}$ by 
means of the Operator Product Expansion. Once $\rho(s)$ is parameterized by
a hadronic representation inferred from experimental data, the resulting
Eq.~(\ref{borpsi}) represents a ($M^2$-dependent) equation relating $m_s$ 
to experimental data and known QCD parameters.

To arrive at the final result, one applies the notion of global quark-hadron 
duality, which essentially consists in identifying
${\int_{s_0}^\infty ds \rho(s) e^{-s/M^2}}\cong 
{\int_{s_0}^\infty ds \rho_{OPE}(s) e^{-s/M^2}}$, 
 since, due to asymptotic freedom, quark and gluon degrees 
of freedom (rather than hadrons) should dominate 
above an effective energy 
threshold $s_0$. Accordingly, 
Eq.~(\ref{borpsi}) takes the form
\begin{equation}
\Psi^{''}(M^2)|_{OPE}={1 \over M^6} 
\int_{0}^{s_0} ds \; \rho(s) \; e^{-s/M^2} +
{1 \over M^6} \int_{s_0}^\infty ds \; \rho_{\;OPE}(s) \; e^{-s/M^2} \;\;\; , 
\label{psiope} 
\end{equation}
and the ultimate numerical determination of $m_s$ will be the one for 
which this relation is stable in the, as yet undetermined, parameters 
$M^2$ and $s_0$. 

The OPE expression for $\Psi^{''}(Q^2)$ can be given in terms of a perturbative 
and a non perturbative contribution:
\begin{equation}
\Psi^{''}(Q^2)=\Psi^{''}_P(Q^2)+\Psi^{''}_{NP}(Q^2) \;\;\;. \label{dec}
\end{equation}

For illustrative purposes, we report here just the 
leading order expression of $\Psi^{''}_P(Q^2)$:
\begin{eqnarray}
\Psi^{''}_P(Q^2) &=& {3 \over 8 \pi^2} 
{({\overline m_s}(\mu)-{\overline m_u}(\mu))^2 \over Q^2} 
\Big\{1 + 
{11 \over 3} {\alpha_s (\mu) \over \pi}-2 {\alpha_s (\mu) \over \pi} \log{Q^2 
\over \mu^2} \Big\} \nonumber \\
&-& {6 \over 8 \pi^2}  {
{\overline m_s}^2(\mu) 
({\overline m_s}(\mu)-{\overline m_u}(\mu))^2 \over Q^4} 
\Big\{1 + 
{28 \over 3} {\alpha_s (\mu) \over \pi}-4 {\alpha_s (\mu) \over \pi} \log{Q^2 
\over \mu^2} \Big\}  \; \;\; , 
\label{psipert} 
\end{eqnarray}
and we refer to \cite{chetyrkin} for the explicit, lengthy expressions 
of the order $\alpha_s^2$ and $\alpha_s^3$ contributions.\footnote{
We keep $m_u \ne 0$ only in the coefficient of the perturbative function
(\ref{psipert}).}
In 
Eq.~(\ref{psipert}), $\mu$ is an {\it a priori} arbitrary renormalization 
mass scale. Since $\Psi^{''}(Q^2)$ is related to a physical observable, it 
obeys the homogeneous renormalization group equation:
\begin{equation}
\mu {d \over d \mu}\Psi^{''}(Q^2)=0 \;\;\;  \;\;, \label{rqe} 
\end{equation}
and therefore the scale dependence of the renormalized parameters $\alpha_s$ 
and  $m_s$ appearing in the perturbative calculation of $\Psi^{''}(Q^2)$ 
must cancel against $\log \mu$ factors also appearing in $\Psi^{''}(Q^2)$.
The Borel transform of $\Psi^{''}(Q^2)$ in the approximation of 
Eq.~(\ref{psipert}) is given by:
\begin{eqnarray}
\Psi^{''}_P(M^2) = 
{3 \over 8 \pi^2} 
{({\overline m_s}(\mu)-{\overline m_u}(\mu))^2 \over M^2} \Big\{1 + 
{11 \over 3} {\alpha_s (\mu) \over \pi}-2 {\alpha_s (\mu) \over \pi} 
\Big(\log{M^2 \over \mu^2}+\psi(1) \Big) \Big\} && \nonumber \\
- {6 \over 8 \pi^2}  {{\overline m_s}^2(\mu) 
({\overline m_s}(\mu)-{\overline m_u}(\mu))^2\over M^4} \Big\{1 + 
{28 \over 3} {\alpha_s (\mu) \over \pi}-4 {\alpha_s (\mu) \over \pi} 
\Big(\log{M^2 \over \mu^2}+\psi(2)\Big) \Big\} && ,  
\label{pertbor} 
\end{eqnarray}
where $\psi(x)$ is the dilogarithmic function. The choice $\mu=M$ allows to 
resum the logarithmic terms, transforming the dependence on $\mu$ of the 
running mass ${\overline m}(\mu)$ and of the running coupling constant 
$\alpha_s(\mu)$ into a dependence on the Borel parameter $M$.

In an analogous way one can write the non perturbative contribution to 
$\Psi^{''}(Q^2)$. Referring to \cite{jamin} for the detailed expression 
of the operator expansion, we report here only the Borel transformed 
contribution of $D=4$ operators:
\begin{eqnarray}
\Psi^{''}_{NP}(M^2)&=&
{({\overline m_s}-{\overline m_u})^2 \over M^6} 
\Big\{2 <m_s \bar u u>_0 \Big(1 + {\alpha_s \over\pi} 
({14 \over 3}- 2 \psi(1) -2 \log{M^2 \over \mu^2}) \Big ) \nonumber \\
&-& {I_G \over 9} \Big( 1+ {\alpha_s \over\pi} ({67 \over 18} - 2 \psi(1) 
-2 \log{M^2 \over \mu^2}) \Big)
+ I_s  \Big( 1+ {\alpha_s \over\pi} ({37 \over 9} - 2 \psi(1) 
-2 \log{M^2 \over \mu^2}) \Big) \nonumber \\
&-& {3\over 7 \pi^2} {\overline m_s^4} 
\Big( {\pi \over \alpha_s} + {5\over 6} - {15\over 
4}\psi(1)
-{15\over 4} \log{M^2 \over \mu^2} \Big) \Big\} \;\;\;.
\label{nonpertbor} 
\end{eqnarray}
$I_s$ and $I_G$ are RG invariant combination given by (for $n_f=3$):
\begin{eqnarray}
I_s&=&m_s <\bar s s>_0 + {3 \over 7 \pi^2} m_s^4 ({\pi \over \alpha_s}
- {53 \over 23}) \noindent \\
I_G&=&
-{9\over 4} <{\alpha_s\over \pi} G^2> (1+ {16\over 9} {\alpha_s\over \pi})
+ {4 \alpha_s \over \pi} (1+ {91\over 24} {\alpha_s\over \pi}) m_s <\bar s s>_0
\nonumber \\ 
&+& {3\over 4\pi^2} (1+ {4\over 3} {\alpha_s\over \pi}) m_s^4 \;\;\;.
\end{eqnarray} 
The hadronic contribution to the spectral function $\rho(s)$ can be obtained 
by inserting a set of intermediate states with $J^P=0^+$ and 
$I={1 \over 2}$ into the correlator (\ref{corr}). The simplest examples 
are the two-particle states $|K \pi>$, $|K \eta>$, 
$|K \eta^\prime>$.
In particular, the contribution of the $|K \pi>$ intermediate state, which 
is expected to be the dominant one and whose features are better known from 
the theoretical as well as the experimental point of view,
\footnote{Multiparticle states should be suppressed by 
phase space.} can be written as:
\begin{equation}
\rho^{(K \pi)}(s)={3 \over 32 \pi^2} {\sqrt{(s-s_+)(s-s_-)} \over s} |d(s)|^2 
\hskip 1 cm (s>s_+) \label{rhokpi}
\end{equation}
\noindent where $s_\pm=(M_K \pm M_\pi)^2$, and $d(s)$ is the 
scalar form factor related to the $K_{\ell 3}$ decay form factors $f_{\pm}$:
\begin{equation}
<\pi^0(p^\prime)|{\bar s}\gamma_\mu u|K^+(p)>={1 \over \sqrt{2}} 
[(p+p^\prime)_\mu f_+ + (p-p^\prime)_\mu f_-] \hskip 1cm 
 \; ,
\end{equation}
\begin{equation}
d(s)=(M_K^2-M_\pi^2)\Big[f_+(s) + 
{s \over M_K^2-M_\pi^2} f_-(s) \Big]=(M_K^2-M_\pi^2) f_0(s) \;\;\; . 
\label{ds}
\end{equation}
\noindent 
From the theoretical point of view, $d(s)$ and $f_0(s)$ can be considered 
as analytic functions of the complex variable $s$, with a cut on the real 
axis starting at the threshold $s_+=(M_K+M_\pi)^2$. 

Furthermore, in the 
range $0 \le s \le s_-$, which is the physical one for $K_{\ell 3}$ decay, 
$f_0(s)$ admits a linear expansion for small $s$:
\begin{equation} 
f_0(s)=f_0(0)\Big(1 +\lambda_0  {s \over M_\pi^2} \Big) \label{linear}
\end{equation}
where, from one-loop chiral perturbation theory \cite{gasser1,daphne}, 
one has the  predictions: 
\begin{equation}
f_0(0)=0.973;\qquad\qquad  \lambda_0=0.017 \pm 0.004  
\;\;\;. \label{l0}
\end{equation}
The theoretical value of the slope $\lambda_0$ in Eq.~(\ref{l0}) 
agrees with the experimental result from the high statistics 
analysis of $K^0_{\mu 3}$ decays, which gives the result 
${\lambda_0=0.019 \pm 0.004}$ \cite{donaldson}.\footnote{
\baselineskip 12pt 
However, this value seems not quite in agreement with the result from
${K^+_{\mu 3}}$ decays \cite{PDG82}.}
The linear extrapolation of Eq.~(\ref{linear}) with the theoretical values 
(\ref{l0}) from the decay region to the threshold $s_+$ would imply the 
prediction $d(s_+)=0.30\pm 0.02$.
\footnote{
\baselineskip 12pt
Actually, a small curvature from higher order terms in the low energy 
representation of $f_0$ is admitted, and in principle might be perceptible 
at $s_+$ which is much larger than the decay endpoint $s_-$.} 

As a final constraint, inspired by the quark counting rules 
\cite{brodsky}, we assume  the asymptotic $s$-behaviour: $d(s) \sim 1/s$.

Experimental information on the $K \pi$ system was obtained from the 
analysis of the reaction $K^- p \to K^- \pi^+ n$ some time ago \cite{aston}. 
The partial wave analysis of $K \pi\to K\pi$  
provides evidence, in the $0^+$, $I=1/2$ channel of a well-established 
$K^*_0(1430)$ resonance, 
with $M_R=1429\pm 4 \pm 5 \; MeV $ and $\Gamma_R=287 \pm 10 \pm 21 \; MeV $,
 and a signal for a not yet confirmed 
$K^*_0(1950)$ state, with $M_R=1945 \pm 10 \pm 20 \; MeV $
 and $\Gamma_R=201 \pm 34 \pm 79 \; MeV $ \cite{aston,pdg}. 
Moreover, a non negligible nonresonant component underlying the 
$K^*_0(1430)$ resonance shows up in the measured low energy $K \pi$ 
phase shifts. 

In \cite{chetyrkin, jamin}, the behaviour of $\vert d(s)\vert^2$ 
appearing in Eq.~(\ref{rhokpi}) is modeled by the sum of two Breit-Wigner 
forms (with masses and widths as reported previously), normalized at 
$s=s_+$ to the theoretical prediction from (\ref{linear}) and (\ref{l0}). 
To model an alternative parameterization of (\ref{rhokpi}) which 
includes all the experimental information mentioned above, in particular 
the existence of the nonresonant component, one can 
attempt a construction of the form factor $d(s)$ based on 
analyticity properties and asymptotic behaviour, with the measured 
$K \pi$ phase shifts as an input, consistently with the final state interaction 
theorem.
Such a construction of $d(s)$ can be realized 
by assuming the following representation 
\cite{omnes}:
\begin{equation}
d(s)=d(0) \; exp \Big[ {s \over \pi}\; {\int}_{s_+}^\infty d s^\prime 
{\delta(s^\prime) \over s^\prime (s^\prime -s-i\epsilon)} \Big] 
\label{omnes}\; ,
\end{equation}
\noindent 
with $d(0)$ determined from Eqs. (\ref{ds}), (\ref{l0}) and $\delta(s)$ the 
$K \pi$ $\big(I={1\over 2},\;J^P=0^+\big)$ phase shift.
As usual in the applications of this representation, we do 
not consider the possibility of zeroes for $d(s)$, which would require also a 
polynomial factor in (\ref{omnes}).

The $K \pi$ phase shift is well known in the range 
of invariant mass from $s_+$, to $(1.7\; GeV)^2$: 
it can be parameterized as the  sum of an effective range formula and a
resonant phase:
\begin{equation}
\delta(s)=\delta_{ER}(s) + \delta_{BW}(s) \label{delta}
\end{equation}
\noindent with: 
\begin{equation}
\delta_{ER}(s)=arctg \; \big[ a q(s) (1+b q^2(s)) \big] \label{deltaer}
\end{equation}
\noindent 
($q(s)$ is the $K \pi$ CM momentum) 
and
\begin{equation}
\delta_{BW}=arctg  \big[ {M_R \Gamma_R(s) \over M^2_R -s} \big]\;\;\;,
\label{deltabw} 
\end{equation}
\noindent where $M_R$ is the mass and
$\Gamma_R(s)$ is the $s$-dependent width of the $K^*_0(1430)$ state:
\begin{equation}
\Gamma_R(s)={M_R \over \sqrt s} {q(s) \over q(M^2_R)} \Gamma_R
\end{equation}

\noindent
The parameters $a$ and $b$ have been fitted,  with the result \cite{jamin}:
\begin{equation}
a=2.06 \; GeV^{-1} \;\; , \hskip 1.5 cm b=-1.37\; GeV^{-2} \;\;\; .
\label{ab} 
\end{equation}
\noindent As for the region $s>(1.7 \; GeV)^2$, above the 
 $K \eta^\prime$ threshold, inelastic effects are observed
\cite{aston,tornqvist}.\footnote{The contribution of the $K \eta$ state is 
flavour-$SU(3)$ suppressed \cite{tornqvist}.}
The inclusion of such effects would require a coupled channel analysis of the 
$K \pi$ and $K \eta^\prime$ states, with further contributions 
to $\rho^{(HAD)}$. Since we do not consider here such additional contributions, 
strictly speaking the result of our analysis is a lower bound for $m_s$,
due to the positivity properties of the spectral function $\rho(s)$, 
However, the exponential factor in Eq.~(\ref{borpsi}) should suppress the 
contribution of higher states. This is confirmed by the numerical analysis of
ref.~\cite{chetyrkin}, where the contribution of the $K^*_0(1950)$ has a very 
small influence on the result for $m_s$.

As far as $d(s)$ is concerned, the asymptotic ${1/s}$  behaviour 
at large $s$ can be obtained from (\ref{omnes}) if
$\delta(s) \to 180^0$, and in this regard we fix $\delta(s)=180^0$ for 
$s > (1.7 \; GeV)^2$.
This is a delicate assumption from the numerical point of view, which, however,
is supported by the available experimental data \cite{tornqvist,aston}.

We can remark that, using the parameterization (\ref{delta})
of the phase shift, the form factor $d(s)$ in (\ref{omnes}) 
reproduces the slope predicted by chiral perturbation theory,
as already observed in \cite{jamin}  and shown in Fig. 1. 
This feature is not obtained, in the framework of the representation 
(\ref{omnes}) for $d(s)$, if the phase shift is parameterized in terms of the 
resonant $\delta_{BW}(s)$ phase only. Therefore, 
in the approach considered here,
the inclusion of the  nonresonant component
 is needed on phenomenological grounds.
The obtained spectral function is depicted in Fig. 2, where
we compare the result from the parameterization 
(\ref{delta})-(\ref{deltabw}) with the case of the pure Breit-Wigner form
obtained using $a=b=0$. 
\footnote{The identity of the representation (\ref{omnes}) with a Breit-Wigner 
form in the case of a pure resonance can be shown directly \cite{barton}.}
The substantial reduction of the resonance peak shows 
that even a moderate nonresonant contribution in the $K \pi$ phase shift 
$\delta(s)$ can generate a significant variation of the spectral function
{\it via} the exponential form of the analytic representation (\ref{omnes}).

At this point, we perform the numerical analysis 
of the sum rule, following the same procedure adopted in \cite{chetyrkin}. 
The only slight difference with respect to \cite{chetyrkin} is that, in the 
expansion of the $\beta$ function relevant to the perturbative part of the sum 
rule:
\begin{equation}
\beta(a)=\sum_n \beta_n a^n \hskip 2 cm \Big( a={\alpha_s \over \pi} \Big) 
\label{beta}
\end{equation}
\noindent we use
the coefficient $\beta_4$ recently computed in the $\overline {MS}$ 
scheme \cite{larin}, namely the set of values: 
\begin{equation}
\beta_1=-{9 \over 2}, \;\;\; \beta_2=-8, \;\;\; \beta_3=-{3863 \over 192},
\;\;\; 
\beta_4=-{ 281198 \over 4608} - {890 \over 32} \zeta(3)
\label{beta4}
\end{equation}
\noindent 
for $N_c=3$ and $n_f=3$; $\zeta$ is the Riemann zeta function.
The numerical value for $\beta_4$ differs by a factor of two from that used in 
\cite{chetyrkin} obtained by a Pad\'e approximant \cite{pade}. 

Moreover, for the analogous expansion of the anomalous dimension 
$\gamma(a)$: $\gamma(a)=\sum_n \gamma_n a^n$, we use the first 
three computed coefficients (for $N_c=3$, $n_f=3$)
\begin{equation}
\gamma_1=2, \;\;\; \gamma_2={91 \over 12}, \;\;\; \gamma_3={8885 \over 288} -5 
\zeta(3) \;\;\;, \label{tregamma}
\end{equation}
\noindent whereas for $\gamma_4$ we use the result of two Pad\'e approximants 
\begin{equation}
{\gamma(a) \over a}={A+Ba \over 1+Qa} \label{pade1}
\end{equation}
($A,B,Q$ numerical coefficients)
giving 
$\gamma_4={\gamma_3^2 /\gamma_2}$ \cite{chetyrkin,kataev}, 
and
\begin{equation}
{\gamma(a) \over a}={A \over 1+Ra+Sa^2}  \label{pade2}
\end{equation}
giving $\gamma_4=2{\gamma_2 \gamma_3 / \gamma_1} - {\gamma_2^3 /\gamma_1^2}$.
The difference between the two approximants 
(\ref{pade1}) and (\ref{pade2}) is about $3 \%$, and has no practical
influence on the final result for $m_s$.
\footnote{After this paper was completed, a calculation of $\gamma_4$ in the 
${\overline {MS}}$ scheme appeared in \cite{chet97}. The computed value:
$\gamma_4=88.5258$ differs from the results in (\ref{pade1},\ref{pade2})
by $12\%$. This effect has no influence on the result for $m_s$.}
As for the parameters in the theoretical side of the sum rule, we use the 
values: $<{\bar u}u>|_{\mu=1\; GeV}=(-0.225\;GeV)^3$, 
$\displaystyle <{\alpha \over \pi}G^2>=(2-6) \; 10^{-2}\; GeV^4$ 
and 
$\displaystyle {<{\bar s}s> \over <{\bar u}u>}=0.7-1.0$; in the coefficient of 
eq. (\ref{psipert})
we use $m_u=m_s/34.2$ as from Eq.~(\ref{cratios}). 
The dependence of the result for $m_s$ from these nonperturbative contributions 
is rather weak, and therefore the uncertainty corresponding to the variation of 
the parameters in the considered ranges is found to be
small, of the order of $1 \; MeV$.

We varied the threshold $s_0$  in the range $5-7\; GeV^2$,
the Borel parameter $M^2$ in the range $1-9 \; GeV^2$, and 
we considered the three values for 
$\Lambda^{n_f=3}_{\overline {MS}}$: $280$, $380$ and $480$ $MeV$.
Specifically,
the best stability in $M^2$ of the resulting $m_s$ is obtained for
$s_0=5.5 \; GeV^2$ when  $\Lambda_{\overline {MS}}^{n_f=3}=380\; MeV$; 
for  $s_0=5.0 \; GeV^2$ 
when $\Lambda_{\overline {MS}}^{n_f=3}=280 \; MeV$ and for $s_0=6 \; 
GeV^2$ when $\Lambda_{\overline {MS}}^{n_f=3}=480\; MeV$. 
\footnote{
Concerning the possible instanton contribution to the correlator 
(\ref{corr}) \cite{istantons}, 
 even though a quantitative assessment is difficult, 
we argue that such contribution is within the uncertainty 
related to the hadronic spectral function, mainly due to the large energy scale 
involved in the calculation of the correlator Eq.~(\ref{corr}).}

The stability curves for the running mass ${\overline m}_s(1 \; GeV)$ are 
depicted in Fig. 3
where only the dependence on $\Lambda_{\overline {MS}}^{n_f=3}$ and $M^2$ is 
displayed. Changing only 
$\Lambda_{\overline {MS}}^{n_f=3}$ the result for $m_s$ changes in the range 
$130-140 \; MeV$. An additional uncertainty is due to the parameters describing 
the continuum $K\pi$ component; changing the parameter $a$ in the effective 
range formula (\ref{deltaer}) 
by $20\%$, and the other parameters in the considered 
ranges, we have
\begin{equation}
{\overline m}_s(1 \; GeV) = 125-160 \; MeV \label{result0}
\end{equation}
with a lower bound
\begin{equation}
{\overline m}_s(1 \; GeV) \ge 120 \; MeV. \label{result}
\end{equation}
\noindent 
The value (\ref{result0}) derives from a parameterization of 
the hadronic spectral function which  has
the correct analyticity properties and uses all the available experimental 
information.
This determination can be systematically improved by a dedicated 
analysis of the scalar $I={1 \over 2}$ $K \pi$ channel using, e.g., 
the semileptonic decays of the $\tau$ lepton.

In principle, the result (\ref{result0}) 
is reflected in the 
lower bound (\ref{result}) due to the neglect of higher states;
however, we do not expect that
the actual determination of $m_s$ in the theoretical framework considered here
can significantly differ from it.
The value (\ref{result0}), lower that the result obtained assuming the 
resonance dominance, shows the significancy of the $K \pi$ nonresonant 
continuum.
\footnote{Similar conclusions have also been reached in ref.\cite{gupta1}.}

\clearpage

\hskip 3 cm {\bf FIGURE CAPTIONS}
\vskip 1 cm
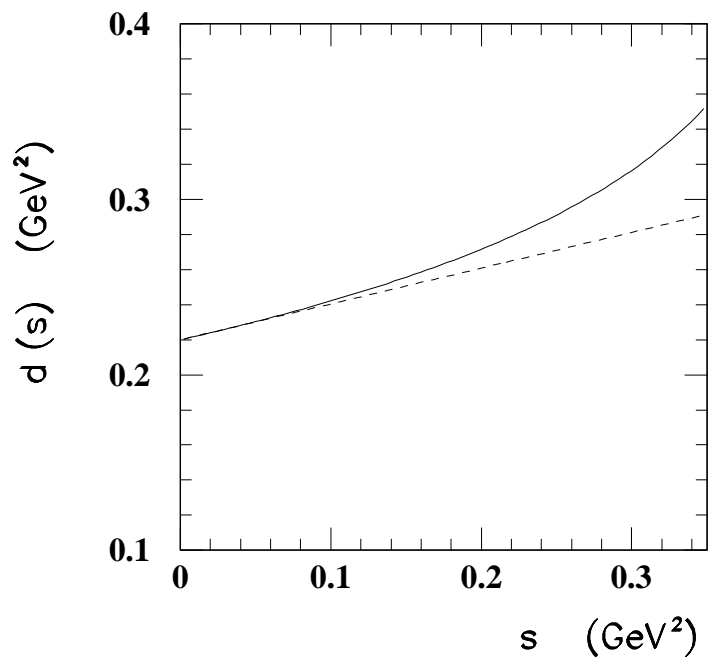
\noindent {\bf Fig. 1}\\
\noindent
Form factor $d(s)$ obtained using the representation Eq.~(\ref{omnes})
(continuous line). The $\chi$PT prediction Eq.~(\ref{linear}), with 
$\lambda_0=1.69 \; 10^{-2}$, is also shown (dashed line).
\vspace{5mm}
\vskip 1 cm
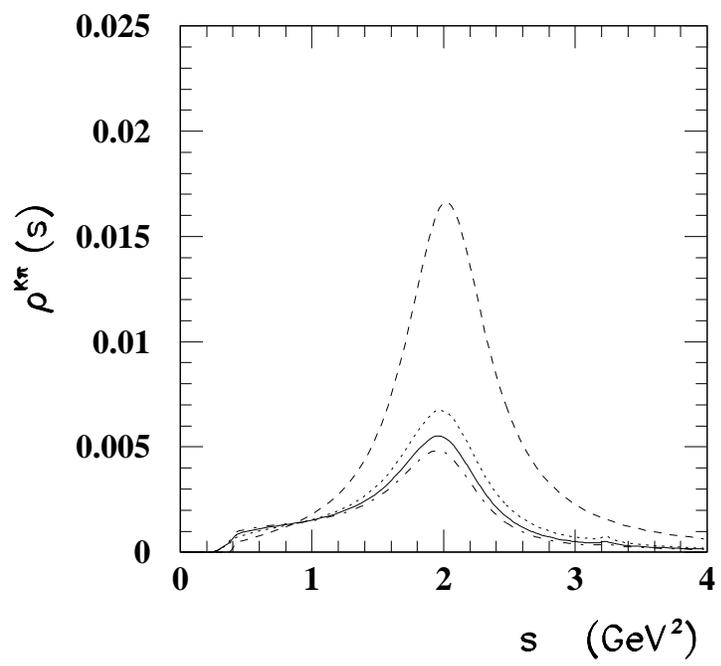
\noindent {\bf Fig. 2}\\
\noindent
The hadronic spectral function $\rho(s)$ obtained 
including only the resonant $K^*_0(1430)$ state (upper dashed line), 
or using the representation Eq.~(\ref{omnes}) (lower lines); the three 
low-lying curves correspond to a $20 \%$ variation of the parameter $a$ in 
Eq.(\ref{deltaer}).
 
\vspace{5mm}
\vskip 1 cm
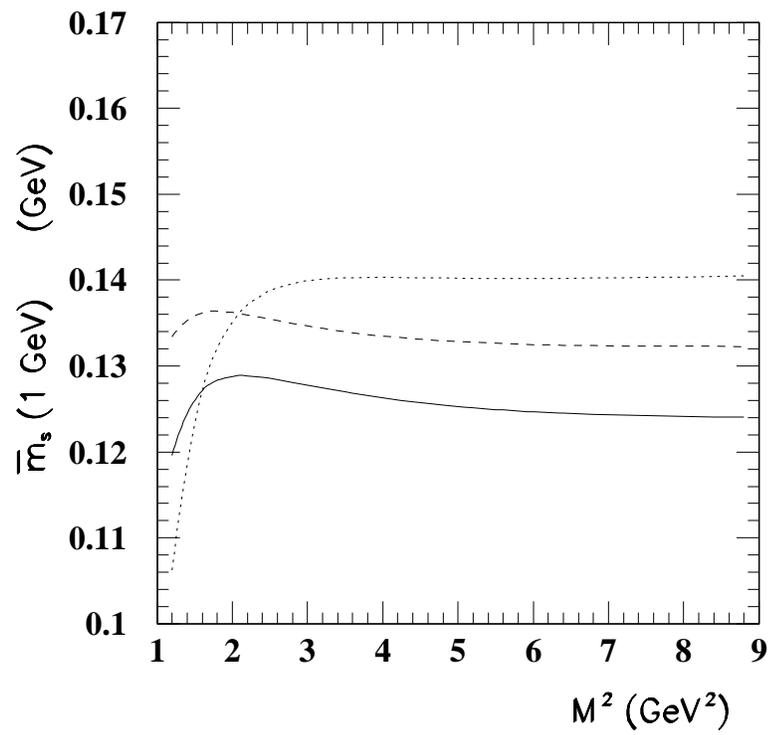
\noindent {\bf Fig. 3}\\
\noindent
The running mass ${\overline m_s}(\mu)$ at the 
scale $\mu=1\;\;GeV$ as a function of the Borel parameter $M^2$, 
for $\Lambda_{\overline {MS}}^{n_f=3}=380\; MeV$ (continuous line),
$\Lambda_{\overline {MS}}^{n_f=3}=280\; MeV$ (dotted line) and 
$\Lambda_{\overline {MS}}^{n_f=3}=480\; MeV$ (dashed line).
\vspace{5mm}
        
\end{document}